\documentclass[aps,prd,10pt,twocolumn,showpacs,preprintnumbers,amsmath,amssymb,nofootinbib,floatfix,hyphens,break]{revtex4-1}

\usepackage{graphicx, bm, color} 
\usepackage[breaklinks]{hyperref}

\usepackage{dcolumn}

\usepackage{textgreek} 
\usepackage{etoolbox} 

\def\twod/{2D(3$v$)}
\def\um/{\textmu m} 
\newcommand{\Wcm}[2][0]{\ifnumcomp{0}{=}{#1}{}{$#1~\cdot$~}$10^{#2}$~W~cm$^{-2}$} 
\def\mildI/{\Wcm[5.4]{17}}
\def\relI/{\Wcm[3]{18}}

\begin{document}







\title{Particle-in-Cell Simulations of Electron Beam Production from Infrared Ultra-intense Laser Interactions}

\author{Gregory K. Ngirmang}
 \email{ngirmang.1@osu.edu}
 \affiliation{\mbox{Department of Physics, The Ohio State University, Columbus, OH, 43210, USA}} 
 \affiliation{Innovative Scientific Solutions, Inc., Dayton, OH, 45459, USA}

\author{Chris Orban}
 \affiliation{\mbox{Department of Physics, The Ohio State University, Columbus, OH, 43210, USA}} 
 \affiliation{Innovative Scientific Solutions, Inc., Dayton, OH, 45459, USA}

\author{Scott Feister}
 \affiliation{Department of Astronomy and Astrophysics, University of Chicago, Chicago, IL, 60637, USA}
\affiliation{\mbox{Department of Physics, The Ohio State University, Columbus, OH, 43210, USA}} 

\author{John T. Morrison}
\affiliation{Innovative Scientific Solutions, Inc., Dayton, OH, 45459, USA}

\author{Enam A. Chowdhury}
\affiliation{\mbox{Department of Physics, The Ohio State University, Columbus, OH, 43210, USA}} 
\affiliation{Intense Energy Solutions, LLC., Plain City, OH, 43064, USA}

\author{W. M. Roquemore}
\affiliation{Air Force Research Laboratory, Dayton, OH, 45433, USA}


\date{\today}

\begin{abstract}
Advances in ultra-intense laser technology are enabling, for the first time, relativistic intensities at mid-infrared (mid-IR) wavelengths. Anticipating further experimental research in this domain, we present high-resolution two dimensional Particle-in-Cell (PIC) simulation results using the Large-Scale Plasma (LSP) code that explore intense mid-IR laser interactions with dense targets. We present the results of thirty PIC simulations over a wide range of intensities ($0.03 < a_0 < 39$) and wavelengths ($\lambda = $780~nm, 3~\um/, and 10~\um/). Earlier studies, limited to $\lambda = $780~nm and $a_0 \sim 1$ \cite{Orban_etal2015,Ngirmang_etal2016}, identified super-ponderomotive electron acceleration in the laser specular direction for normal-incidence laser interactions with dense targets. We extend this research to mid-IR wavelengths and find a more general result that normal-incidence super-ponderomotive electron acceleration occurs provided that the laser intensity is not highly relativistic ($a_0 \lesssim 1$) and that the pre-plasma scale length is similar to or longer than the laser wavelength. Under these conditions, ejected electron angular and energy distributions are similar to expectations from an analytic model used in \cite{Ngirmang_etal2016}. We also find that, for $a_0 \sim 1$, the mid-IR simulations exhibit a classic ponderomotive steepening pattern with multiple peaks in the ion and electron density distribution. Experimental validation of this basic laser-plasma interaction process will be possible in the near future using mid-IR laser technology and interferometry.

\end{abstract}

\maketitle


\section{Introduction}
\label{sec:intro}

While advances in laser technology have allowed ultra-intense laser interactions at near-IR wavelengths to be thoroughly explored, and it is only more recently that ultra-intense laser interactions at mid-IR wavelengths have become experimentally possible \cite{lasermag}. A variety of groups are beginning to examine what may be learned from experiments at these wavelengths and how phenomena observed in the near-IR may scale to longer wavelengths \cite{MURI}. Some of this interest stems from the existence of atmospheric ``windows" in the mid-IR \cite{midIRwindows}, while other groups consider how the longer length scale of mid-IR interactions allows subtle phenomena to be more easily probed \cite{MURIresearch}. Another interesting value of intense mid-IR interactions is in examining the physics of laser damage \cite{Austin_etal2015,Austin_etal2016}.

To the best of our knowledge, despite recent interest in mid-IR ultra-intense laser interactions, the literature has not focused much attention on intense mid-IR laser interactions with dense (i.e. solid or liquid density) targets. These interactions are interesting for a variety of reasons, among them the favorable scaling of the ponderomotive electron energy with laser wavelength ($a_0 \sim \sqrt{I \lambda^2} \sim \lambda$). However, given the complexity of ultra-intense interactions with dense targets, these scaling arguments can only offer an order-of-magnitude expectation for the results of detailed simulations and experiments in this regime. With experimental capabilities still maturing in the mid-IR, the present work offers a simulation survey of energetic electron ejection from mid-IR laser irradiated dense targets.

The work presented here is motivated in part by earlier investigations of normal-incidence ultra-intense laser interactions with liquid targets at the Air Force Research Lab which found much larger than expected conversion efficiencies from laser energy to ejected electron energy \cite{Morrison_etal2015}. These experimental results prompted simulation work by \cite{Orban_etal2015,Ngirmang_etal2016}. \cite{Orban_etal2015} presented \twod/ Particle-in-Cell (PIC) simulations showing significant electron ejection at superponderomotive energies and emphasized that ultra-intense laser interactions at the $\sim 10^{18}$~W~cm$^{-2}$ ($a_0 \sim 1$ for $\lambda \approx 800$~nm) intensities present in the experiment should create strong standing-wave fields near the target. \cite{Ngirmang_etal2016} performed 3D PIC simulations of these targets and provided an analytic model to explain both the energies and angular distribution of ejected electrons. More recently, \cite{Feister_etal2016} have reported direct experimental measurements of the ejected electron energies, confirming the existence of multi-MeV electrons in the experiment and from this reinforcing conclusions that the conversion efficiency in the experiment is large compared to other ultra-intense laser experiments. An interesting question, then, is whether superponderomotive electron ejection occurs even with intense mid-IR laser interactions.

The other motivator for this project is the plan to purchase and upgrade an intense 3~\um/ wavelength laser system at the Air Force Research Laboratory.  The upgraded laser system will be able to produce $\sim1$~mJ scale laser pulses  and a peak intensity near $10^{17}$~W/cm$^2$.  We explore a much wider range of laser energies and intensities in an effort to examine the physics of mid-IR laser interactions with dense targets.

Sec.~\ref{sec:sims} describes our simulation setup. Sec.~\ref{sec:results} describes our results. Finally, Sec.~\ref{sec:summary} provides a summary and conclusions.

\newcolumntype{x}[1]{>{\centering\arraybackslash}m{#1}}
\begin{table*}
    \begin{tabular}{| x{0.7in} | x{1.4in} |
                      x{0.7in}  | x{0.5in}  | 
                      x{0.7in}  | 
                      x{0.6in}  | x{0.7in}  |}
    
    \hline
    Wavelength (\um/) & Intensity Range (W/cm$^2$) &
    Gaussian Radius (\um/) & Pulse FWHM (fs) &
    Scaled Pre-Plasma Scale  (\um/) & 
    Simulation Timestep (fs) & Simulation Spatial Resolution (\um/) \\ \hline
    10 & 1.64$\cdot$10$^{13} - $1.64$\cdot$10$^{18}$ &
    27.5 & 513 &
    19.2 & 
    0.5 & 0.25 \\ \hline
    3 & 6.08$\cdot$10$^{14} - $6.08$\cdot$10$^{19}$ &
    8.5 & 154 &
    5.77 & 
    0.15 & 0.1 \\ \hline
    0.78 & 3.46$\cdot$10$^{16} - $3.46$\cdot$10$^{21}$ &
    0.15 & 40 &
    1.5 & 
    0.1 & 0.033 \\ \hline
    \end{tabular}
    \label{table:wavelengthparams}
    \caption{Summary of the parameters for each simulation with a given incident wavelength laser. As explained in \ref{sec:sims}, the performed simulations had parameters, such as laser pulse duration and gaussian radius, chosen to scale proportionally with the wavelength. Also explained in the referenced section is that simulations with 1.5~\um/ were performed for all wavelengths.}
\end{table*}

\section{Particle-In-Cell Simulations}
\label{sec:sims}

We performed 30 different high-resolution \twod/ PIC simulations with the LSP code \cite{Welch_etal2004}. For all simulations the initial conditions included a liquid-density water slab target with some assumed pre-plasma scale length similar to earlier studies \cite{Orban_etal2015,Ngirmang_etal2016}. In all simulations, a laser is normally incident onto the water slab. We use the following Cartesian coordinate system for these simulations: the positive $x$-axis is the direction of the laser, the $y$-axis is the polarization direction, and $z$-axis is the axis of the water column, which is assumed to be the axis of symmetry in the \twod/ PIC simulations.

The simulations involved a normally incident, spatially Gaussian, sine-squared envelope pulse with 780~nm, 3~\um/, and 10~\um/ wavelengths (denoted $\lambda$). These simulations extend the results of earlier investigations with 780~nm laser pulses \cite{Orban_etal2015,Ngirmang_etal2016} by examining the same phenomena with long infrared(IR) wavelengths. For convenience we will often refer to the set of all simulations performed with a particular laser wavelength incident by saying, for example, ``the 3~\um/ simulations'', and likewise for the other wavelengths. All the 780~nm simulations had a laser pulse with a 2.15~\um/ Gaussian radius and a 40~fs temporal full-width-half-maximum (FWHM) pulse duration (similar to the laser system described in \cite{Morrison_etal2015, Orban_etal2015,Ngirmang_etal2016}). The 3~\um/ simulations have laser pulses with 8.25~\um/ Gaussian radius and 158~fs FWHM pulse duration, and the 10~\um/ simulations had laser pulses with 27.5~\um/ Gaussian radii and 510~fs. These Gaussian radii and pulse durations were chosen so that the ratio of the wavelength to Gaussian radius and the number of optical periods in a pulse were fixed across all simulations regardless of laser wavelength. For each wavelength we simulated a range of pulse energies from $10^{-4}$ J, 1~mJ , 10~mJ, 1~J, to 10~J. Since the Gaussian radius and pulse duration were fixed for each wavelength this was done by changing the intensity. Simulation parameters for each wavelength simulation are summarized in Table \ref{table:wavelengthparams}.

The target in all simulations consisted of free electrons, protons, and O$^+$ ions, with number densities set in relative proportion to make the target match water's chemical composition and to ensure charge neutrality (O$^+$ to p$^+$ to  e$^-$ ratio of 1:2:3). In all simulations, the plasma density only varied along the $x$ direction. In the ``target'' region the density is constant and in the ``pre-plasma'' region the density profile is exponentially decreasing in $x$ away from the target region.

For every intensity and wavelength considered we perform a simulation with a 1.5~\um/ scale-length, for direct comparison with previous studies \cite{Orban_etal2015,Ngirmang_etal2016} which employed a such a scale length pre-plasma. For 3~\um/ and 10~\um/ wavelengths, we also perform simulations where the pre-plasma scale length is a constant multiple of the wavelength ($L = 1.92 \lambda$), so that we performed 3~\um/ simulations with a 5.77~\um/ scale length and 10~\um/ simulations with a 19.2~\um/ scale-length. We refer to these scale lengths are the ``scaled'' scale lengths. The 780~nm and 3~\um/ simulations had a target region electron density of $10^{23}$ cm$^{-3}$, which with the O$^+$ to p$^+$ to  e$^-$ ratio mentioned earlier correspondent to the mass density of $1$~g/cm$^{3}$ as one would expect for liquid water. The 10~\um/ simulations with a 1.5~\um/ scale length also had a target region with this same $10^{23}$~cm$^{-3}$ electron density into the target. However, the 10~\um/ simulations with a 19.2~\um/ scale-length had a target region electron density of $10^{21}$ cm$^{-3}$ in order to reduce the size of the target in order to reduce computational requirements. The 780~nm simulations used a target that was 20~\um/ wide in the $y$ direction, the 3~\um/ simulations used a 100~\um/ wide target, and the 10~\um/ simulations used a 220~\um/ wide target. All targets had a initial temperature of 1 eV. All simulations had absorbing boundaries with 10~\um/ between the initial target and the simulation boundaries. 

The classical formula for the critical electron density is $n_c = 4\pi m_e \omega^2/e^2 = m_e / \pi \lambda^2 e^2$. For the 780~nm, 3~\um/, and 10~\um/ wavelengths, this corresponds to $n_c=$1.74$\times$10$^{21}$ cm$^{-3}$, 1.24$\times$10$^{20}$~cm$^{-3}$, and 1.11$\times$10$^{19}$~cm$^{-3}$ respectively. In all simulations, the laser focus was chosen to coincide with the critical density in the pre-plasma layer.
For the 780~nm simulations, a spatial resolution with spacing of 33~nm (roughly 23 cells per wavelength) was used and timesteps of 0.1~fs was used. For the 3~\um/ simulations a spacing of 100~nm and temporal resolution of 0.15~fs was used. Finally, the 10~\um/ simulations utilized a spatial resolution of 250~nm and temporal resolution of 0.5~fs. Although these simulations do not resolve the Debye length in every cell (since there are cells with near-solid densities with sub-nanometer Debye lengths), the phenomena of interest is electron acceleration in the underdense pre-plasma extending from the target where the Debye length is much larger and more easily resolved. The implicit algorithm in LSP avoids grid-heating issues associated with the Debeye instability so that the behavior of near-solid density regions in the simulation does not ruin the overall energy conservation of the simulation.  All simulations had 27 macro-particles per cell per species (free electrons, protons, and O$^+$ ions). As in earlier work \citet{Orban_etal2015}, the O$^+$ ions in this simulation can be further ionized by strong electric fields according to the Ammisov-Delone-Krainov rate \cite{ADK}. In the simulation electron macroparticles scatter by a Monte-Carlo algorithm as in \citet{Kemp_etal2004} with a scattering rate determined by a Spitzer model \cite{Spitzer1963} except at low temperatures where the scattering rate is limited by the timestep. The 1.5~\um/ scale-length simulations were run for three times the duration of the simulated laser pulse (i.e. three times the full-width full max duration of the pulse). The scaled simulations were run for 3.5 times the duration of the simulated laser pulse because these were larger targets with a more extended pre-plasma.

The number of pulse energies investigated, the various scale lengths assumed with the three laser wavelengths add up to a total of 30 \twod/ simulations. The parameters of all simulations are listed in the appendix, and parameters common across given wavelength simulations are summarized in Table ~\ref{table:wavelengthparams}. 

\section{Results}
\label{sec:results}

\subsection{Ejected Electron Energies}
\label{subsec:energies}
\begin{figure*}
\includegraphics[angle=0,width=3.7in]{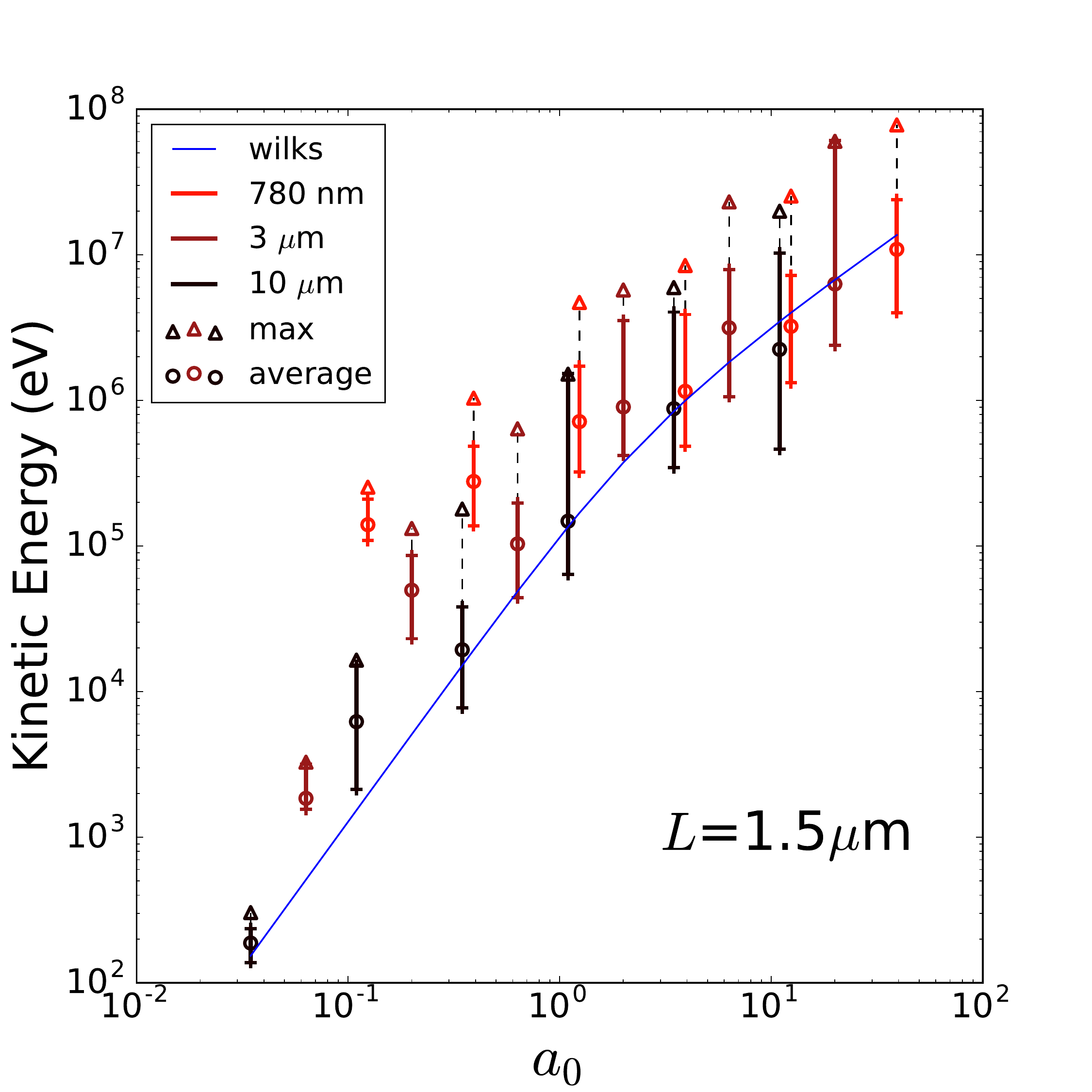}\includegraphics[angle=0,width=3.7in]{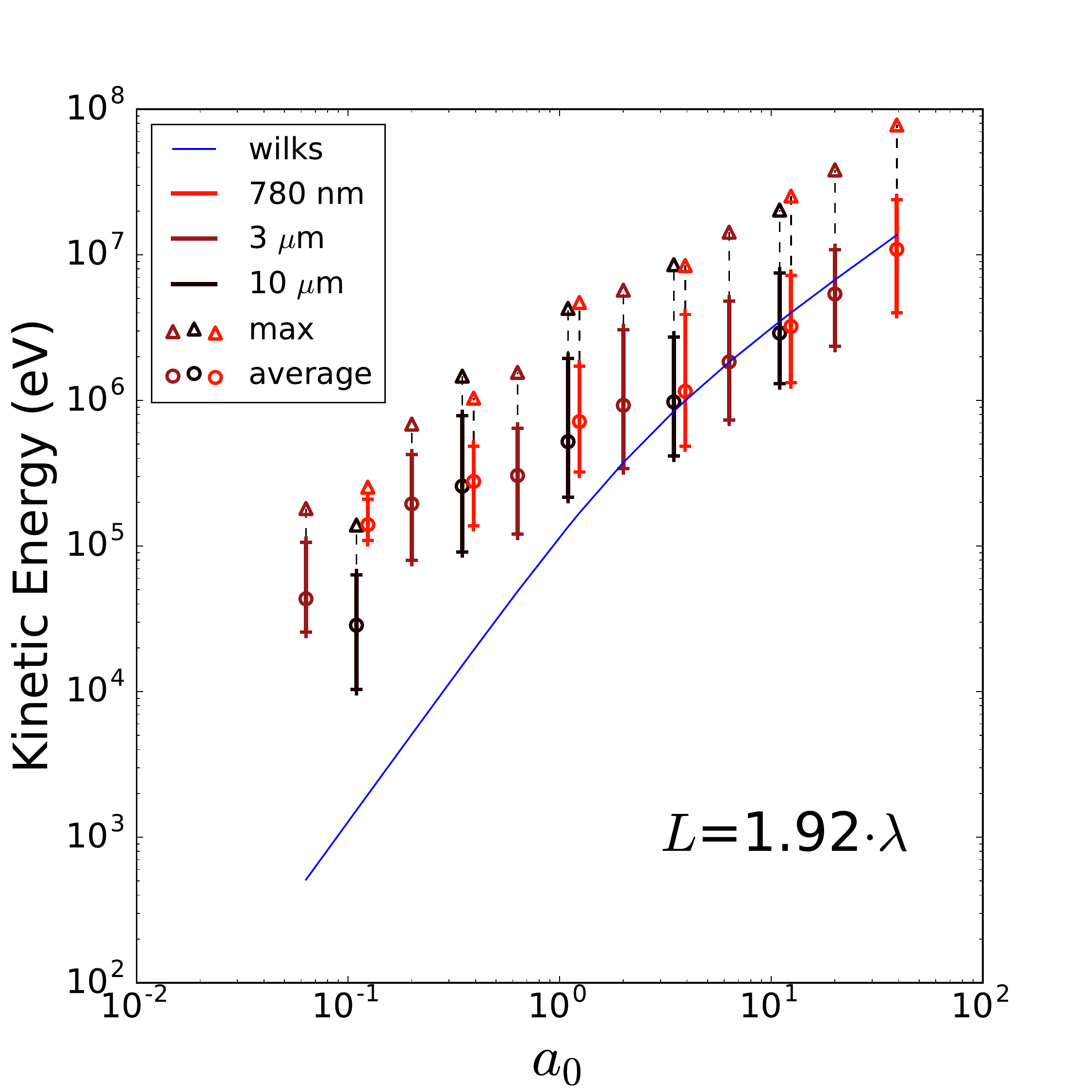}
\vspace{-0.4cm}
\caption{Left panel: Ejected electron energies observed in PIC simulations with laser wavelengths 780~nm (red), 3~\um/ (dark red), and 10~\um/ (black) and with a 1.5 \um/ pre-plasma scale length. Right panel: Ejected electron measurements from simulations where the pre-plasma scale length is "scaled" to the wavelength of the laser such that $L = 1.92 \cdot \lambda$. In both panels results are shown as a function of the $a_0$ of the incident laser. Expectations from Wilks scaling (Eq.~\ref{eq:wilksdef}) are shown as a solid blue line. The lowest $a_0$ case for $L = 1.92 \cdot \lambda$ is not shown as accelerated electrons were not detected in that case. Open circles represent the mean kinetic energy of accelerated electron that reach the edge of the simulation space in a 150 degree apex cone in the backwards direction. The error bars represent 68\% of the electrons around the mean energy bin. The triangles represent the maximum ejected electron energy observed. Dashed lines connect the triangles to the error bars so that it is easy to see that these measurements come from the same simulation.} \label{fig:energies}
\end{figure*}

Fig.~\ref{fig:energies} shows the energy spectra of back-accelerated electrons for all the simulations performed. The left panel of Fig.~\ref{fig:energies} shows the results of the simulations with a 1.5~\um/ scale-length pre-plasma, while the right panel plots the results of the scaled scale-length simulations where the scale-length is proportional to the wavelength ($L=1.92\lambda$). These figures plot the mean and maximum energy of escaping electrons on the $y$-axis as a function of the normalized vector potential $a_0$ of the incident laser. Here, $a_0 = e E_0 / m_e \omega c$, with the peak electric field value of the incident pulse is denoted $E_0$, $c$ is the speed of light, $m_e$ is the mass of an electron, and $\omega = 2 \pi c /\lambda$ is the angular frequency of the laser beam.  The mean electron energy is determined by taking the average energy of electrons that reach the boundary of the simulation. Since we are concerned here with back directed electrons only those electrons with a momentum angle within $\pm75^\circ$ of the incident laser are counted.

These results are found to scale with the Wilks scaling estimate from \cite{Wilks_etal1992},
\begin{equation}\label{eq:wilksdef}
E_{\rm wilks} = \left[ \sqrt{1+\frac{1}{2}a_0^2}-1\right] m_e c^2.
\end{equation}
While there are a number of other formulae that describe the typical energy of electrons interacting with an intense laser field, we choose to compare with Wilks scaling because it is an analytically motivated formula that is reasonably representative of the various scaling models in this regime \cite{Beg_etal1997}. The Wilks model also reduces to the classical ponderomotive energy of an electron in a monochromatic plane wave in the low $a_0$ limit. A binomial approximation yields
\[ E_{\rm wilks} \approx \frac{1}{4} m_ec^2 a_0^2   \]
for small values of $a_0$. This is why, on Fig.\ref{fig:energies}, one sees an $a_0^2$ dependence for Eq.~\ref{eq:wilksdef} at low $a_0$ that transitions to linear dependence (i.e. $\sim a_0$) for $a_0 \gtrsim 1$.

Comparing Eq.~\ref{eq:wilksdef} to the simulation data in the left panel of Fig.~\ref{fig:energies} yields an interesting result that the 10~\um/ wavelength simulations fall closest to the Wilks scaling model prediction. The 3~\um/ and 780~nm wavelength simulations lie significantly above the prediction, especially for low $a_0$ values. The 780~nm wavelength simulations have the most energetic electrons, exceeding Eq.~\ref{eq:wilksdef} by 1-2 orders of magnitude. Thus we say that the ejected electrons in the 780~nm simulations are highly "superponderomotive".

As identified earlier, the right panel of Fig.~\ref{fig:energies} shows the results of the "scaled" simulations. The 780~nm simulations with 1.5~\um/ scale-length appear in both panels of Fig.~\ref{fig:energies}, but the longer wavelength simulations shown in the right panel of Fig.~\ref{fig:energies} all assume a longer scale length than in the left panel. Remarkably, in the right panel the results from all three wavelengths seem to follow roughly the same trend and exceed the Wilks prediction by 1-2 orders of magnitude.

\begin{figure*}
    \centering
    \includegraphics[angle=0,width=7in]{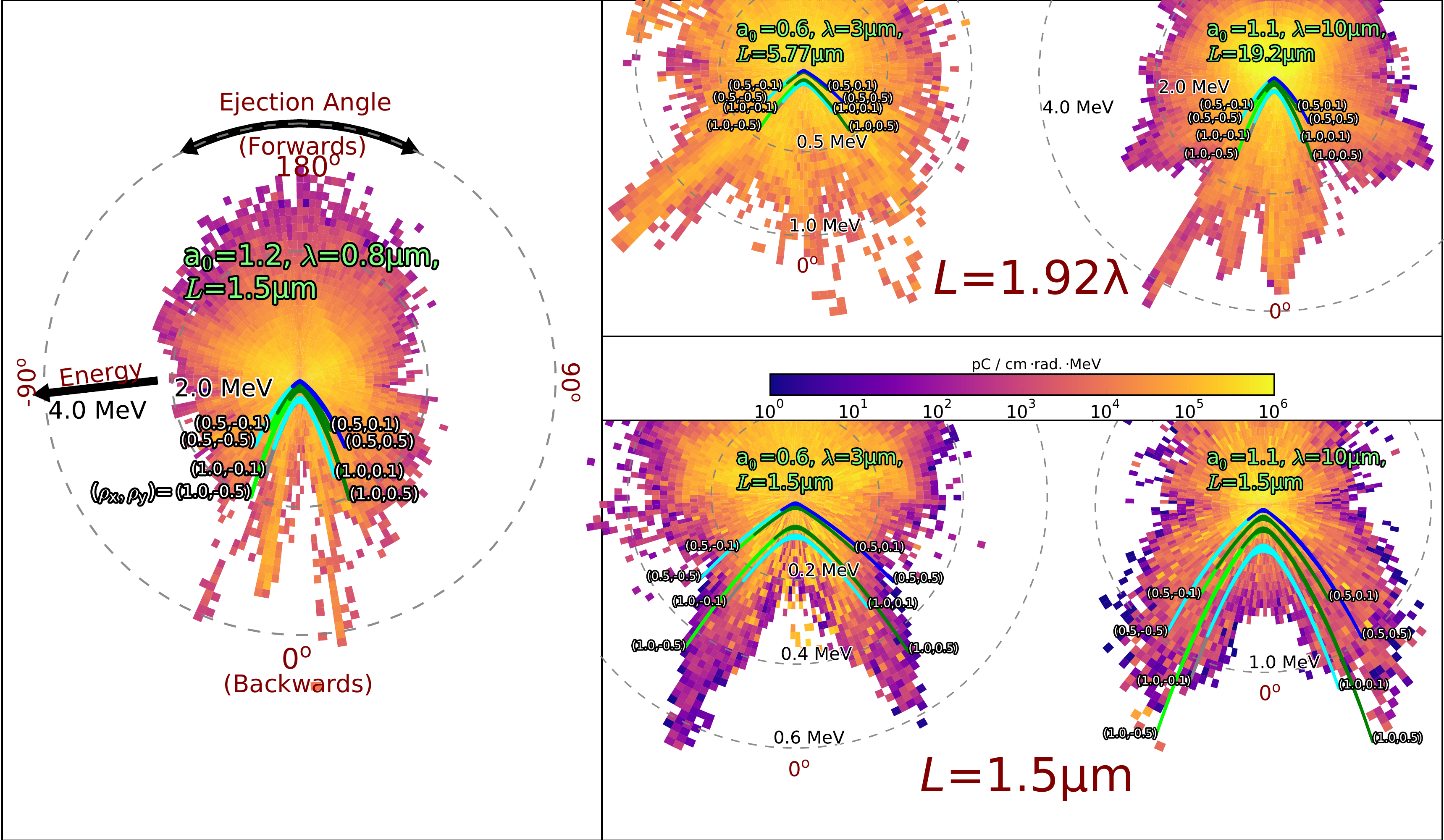}
    \caption{Summary of the kinetic energies and ejection angles of electrons that escape the simulation space of selected PIC simulations in which the incident laser has $a_0\sim1$. In all panels, the distance from the center represents the kinetic energy of escaping electrons while the angle to the origin represents the final ejection angle, and the color is proportional to the amount of charge in a particular energy and angular bin. The left panel shows a 780~nm simulation with $a_0=1.2$ while the two simulations in the center column show two different 3~\um/ simulations with $a_0 = 0.6$ and the right column shows two different 10~\um/ simulations with $a_0 = 1.1$. The upper right panel shows two mid-IR simulations with a $L = 1.92 \lambda$ exponential scale length, while the lower right panel shows two mid-IR simulations with $L = 1.5$~\um/. An analytic model from \cite{Ngirmang_etal2016} is overlayed in each panel and labeled by the normalized momentum similar to figures shown in \cite{Ngirmang_etal2016}.
}
    \label{fig:midIR}
\end{figure*}

Fig.~\ref{fig:midIR} presents detailed information on the energies and ejection angles of electrons that leave the simulation volume. Fig.~\ref{fig:midIR} shows results from the three different wavelengths, highlighting intensities with $a_0 \sim 1$ and the mid-IR simulations with $L = 1.92 \lambda$. We also overplot with solid lines the results from an analytic model described in \cite{Ngirmang_etal2016}. This model considers that the back-directed electrons are ejected at high speed into a pulsed plane wave that approximates the reflected laser pulse. Because of the similarity to earlier work in \cite{Ngirmang_etal2016}, it is unsurprising that the model compares favorably to the 780~nm results shown in Fig.~\ref{fig:midIR}. What is more remarkable is that the model predictions compare similarly well to the mid-IR simulations. 


\subsection{Electron density profiles}
\label{subsec:reflection}
\begin{figure*}[t]
  \includegraphics[angle=0,width=6in]{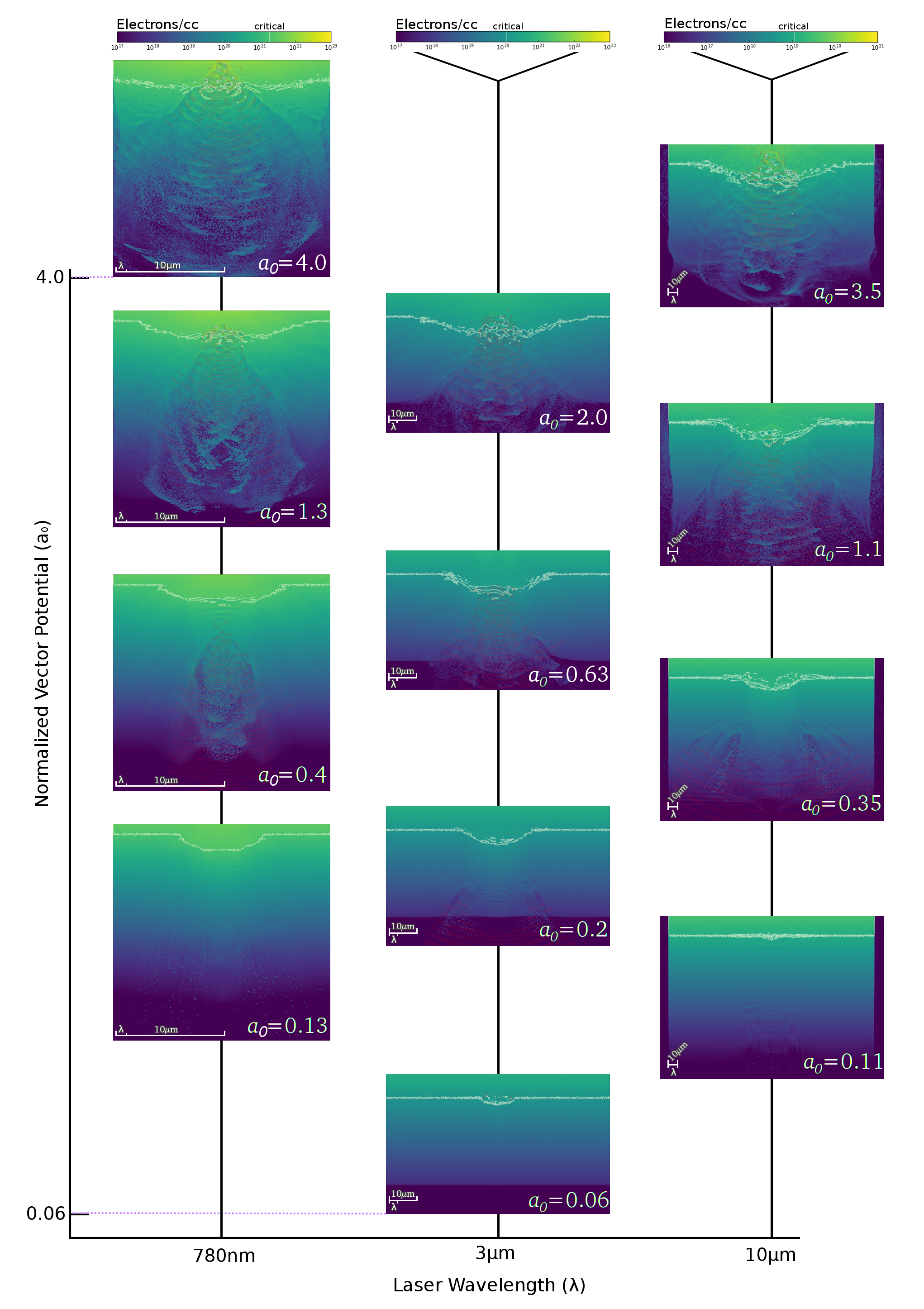}
  \vspace{-0.4cm}
  \caption{
  Electron density snapshots (colormapped rectangles) from twelve unique simulations ranging over wavelengths and $a_0$ values. The snapshots are arranged on the grid according to wavelength and $a_0$ value. Left, middle, and center columns classify the snapshots by the wavelength of the incident laser as 780~nm, 3~\um/, and 10~\um/, respectively. The vertical axis is the $a_0$ value of the incident laser, with the bottom edge of a snapshot aligned along the vertical axis with the peak $a_0$ of the incident laser in that simulation. All colormaps are logarithmic, and roughly chosen normalized to the critical density. Colorbars (top) and spatial scale bars (within each snapshot) are consistent for each column. The snapshots demonstrate two main trends observed across simulations performed, as discussed in Section~\ref{sec:discuss}, that the onset of plasma phenomenon is determined by the $a_0$ value of the incident pulse (as observed scanning up the y-axis) while the size scale of these plasma phenomenon scales with the wavelength (as observed scanning across the x-axis).
  }
  \label{fig:reflection}
\end{figure*}

Fig.~\ref{fig:reflection} provides snapshots of the electron density in the simulation after the reflection of the laser (not more than two pulse-lengths in time later) but while electrons are still moving away from the target. The figure demonstrates that the onset of plasma wave phenomena is dependent on the $a_0$ value of the incident laser pulse, and not only the laser wavelength or laser intensity independently. 
For $a_0<0.2$ for all wavelengths, the escaping electrons are not spatially bunched, and the pre-plasma layer is perturbed less relative to the more intense cases. With $a_0$ values between 0.2 and 0.4, the backwards accelerated electrons break apart the pre-plasma layer as they escape. At $a_0$ values near $\sim0.5$ and above, ejected electrons exhibit bunching, a pattern that becomes more pronounced as $a_0$ becomes larger. As discussed in \cite{Orban_etal2015}, this arises because electrons are only deflected away from the target during two specific moments during the laser cycle. In these simulations, it has been observed that the onset of these bunches is preceded in the process of reflection by a significant electron density hole created by the ponderomotive force, which manifest only as $a_0$ nears and exceeds unity. As $a_0$ exceeds 2.0, as shown in the top two plots, the laser begins to penetrate beyond the (non-relativistic) critical density surface due to relativistic transparency. Hole-boring \cite{Gong_etal2016} does not occur in this case due to the ultrashort pulse in all cases, and the laser pulse begins to penetrate only when the last few cycles of the main pulse are present on the non-relativistic critical density surface.



To comment on another aspect of the plots in Fig.~\ref{fig:reflection}, in essentially all of the plots shown ($0.06 < a_0 < 4$) the laser ionizes the target, moving the critical density (white contour) towards the incoming laser, especially along the laser axis. Because the laser intensity decreases away from the laser axis according to a gaussian spatial profile, this causes the critical density to assume a curved shape as seen in the figure.

\subsection{Ponderomotive Steepening}
\begin{figure*}[t]
\includegraphics[angle=0,width=2.7in]{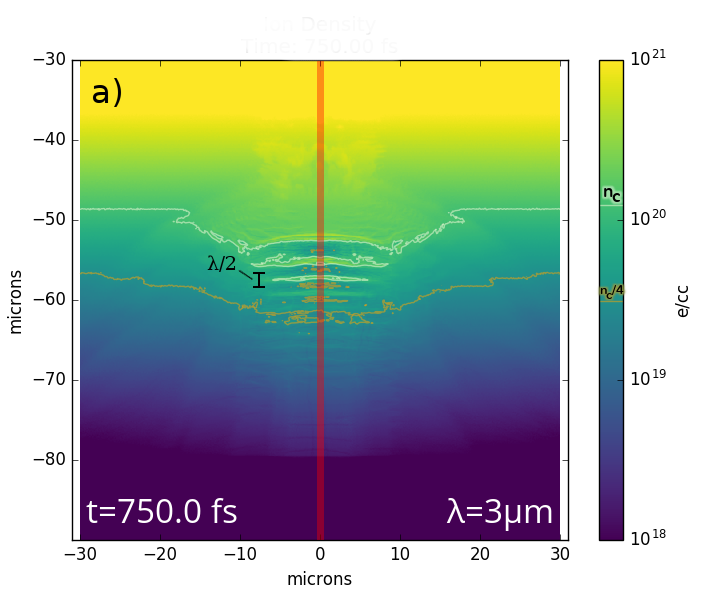}\includegraphics[angle=0,width=3in]{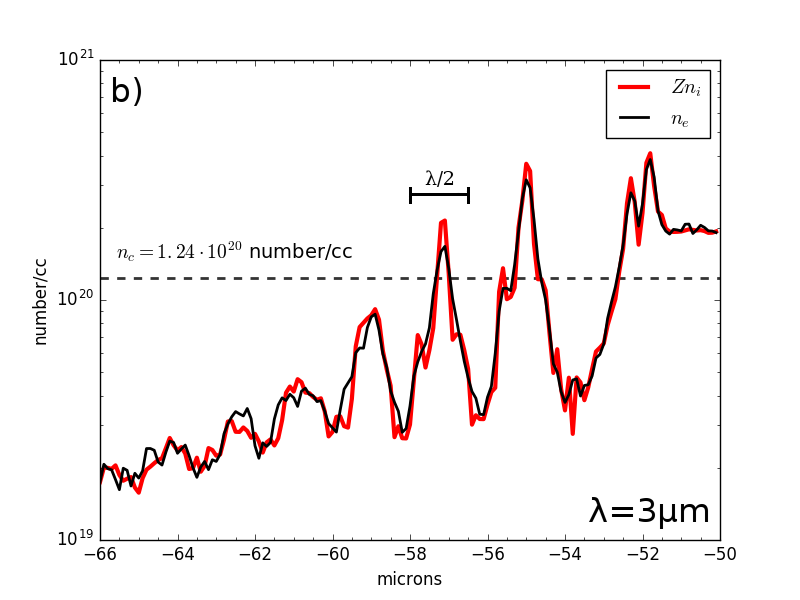}
\includegraphics[angle=0,width=2.7in]{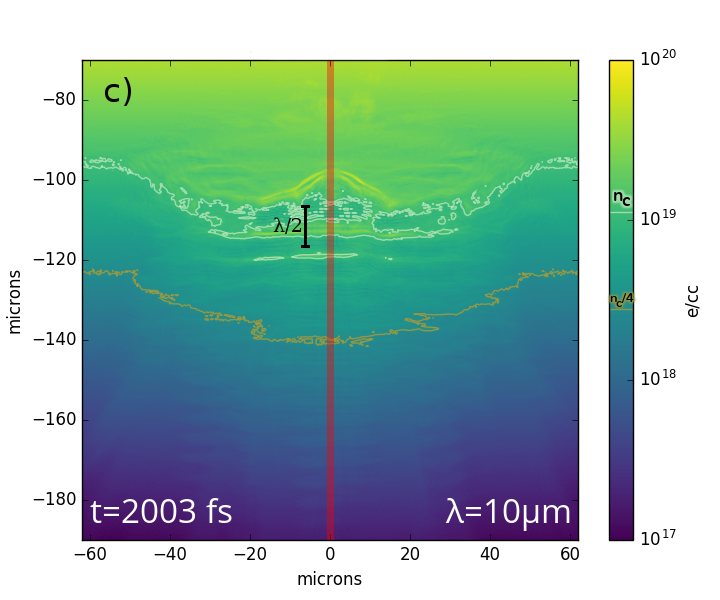}\includegraphics[angle=0,width=3in]{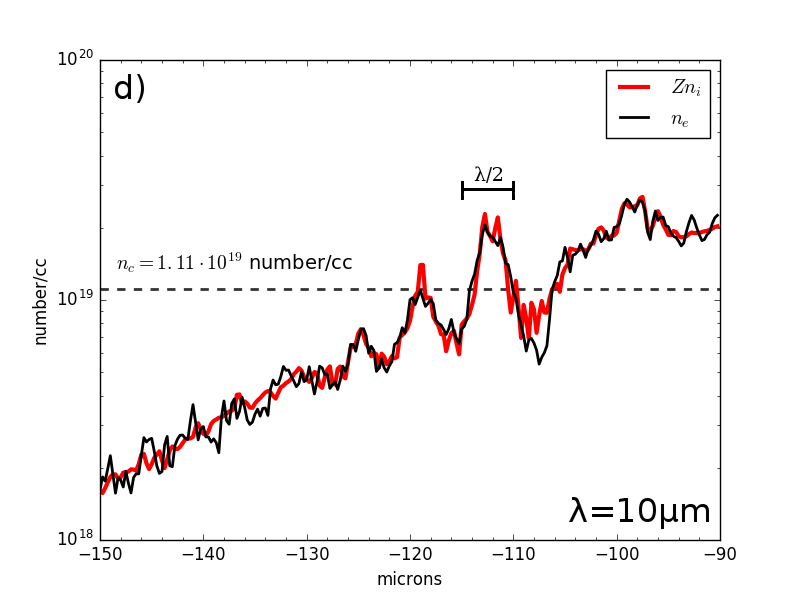}
  \vspace{-0.4cm}
  \caption{ Panel (a) shows a 2D pseudo-color of ion charge density, i.e., $Z n_I $ where $Z$ is the mean charge state of the plasma and $n_I$ is the ion density, in the sub-critical pre-plasma region for a 3~\um/ wavelength simulation with a $6.08\cdot10^{16}$~W/cm$^2$ laser pulse ($a_0 = 0.6$). The pulse is incident on a "scaled" pre-plasma target ($L=$~5.77~\um/). Panel (b) shows a central line out of the ion charge density (red) and electron density (black), with the lineout taken from the highlighted red line in (a). The bottom row (panels (c) and (d)) are similar plots for a 10~\um/ wavelength simulation with a $1.64\cdot10^{16}$~W/cm$^2$ laser pulse. The displayed snapshots shows the target state well after the laser reflection has left the simulation space. On panels (a) and (c), the charge density equivalent to critical electron density is highlighted with white contours and the quarter of critical electron density is highlighted with an orange contour. In all panels, the spatial scale of $\lambda / 2 $ is labeled, given its relevance to ponderomotive steepening as described in \cite{Estabrook_Kruer1983}. }
  \label{fig:ponderomotive}
\end{figure*}

While fig.~\ref{fig:reflection} demonstrates that the onset of plasma wave phenomena is determined by the $a_0$ value of the incident laser, it also demonstrates that at a given $a_0$, plasma features observed scale in physical size with the incident laser wavelength. This finding can be used to scale the physical size the laser plasma interaction to facilitate experimental observation of phenomena which would not be observable with short wavelength pulses. One such phenomenon is ponderomotive steepening, a well-known laser-plasma interaction process where the radiation pressure from the laser modifies the electron density profile, which, over time, will substantially modify the \emph{ion} density profile \cite{Estabrook_Kruer1983} on the scale of half the wavelength. A related process is "hole boring" which has been studied theoretically and experimentally \cite[e.g.][and references therein]{Gong_etal2016}. Unlike hole boring, ponderomotive steepening, as originally described by \cite{Estabrook_Kruer1983}, involves a series of peaks in the ion density profile. Ponderomotive steepening has been observed in a number of PIC simulations in the literature \cite[e.g.][]{Orban_etal2015} but to the best of our knowledge it has never been experimentally observed given the size of such features would be on the sub micron scale. To be able to see this phenomenon, one needs an intense laser and high spatial resolution interferometry of the target that can resolve half-wavelength-scale features. Mid-IR lasers therefore significantly relax the spatial resolution requirements of such an experiment compared to lasers in the near-IR. 

With this in mind we looked for ponderomotive steepening in our simulations and found that both the 3~\um/ wavelength simulation with $a_0 = 0.6$ and $L = 5.77$~\um/ and the 10~\um/ wavelength simulation with $a_0 = 1.09$ and $L = 19.2$~\um/ do exhibit a classic ponderomotive steepening pattern with multiple peaks in both the ion and electron density as shown in Fig.~\ref{fig:ponderomotive}. Ponderomotive steepening was noticed earlier in 800~nm wavelength simulations presented in \cite[][c.f. Fig.12]{Orban_etal2015}. Ponderomotive steepening was also present in simulations that we published in \cite{Ngirmang_etal2016}, although for brevity we did not highlight this result. Fig.~\ref{fig:ponderomotive} provides essentially the first compelling evidence that this effect should persist in mid-IR experiments of this kind.

\section{Discussion}
\label{sec:discuss}

We present 30 high-resolution PIC simulations, which comprise a parameter scan over laser wavelengths, intensities, and target scale-lengths, designed to explore the physics of intense, normally-incident near-IR (780~nm wavelength) and mid-IR (3~\um/ and 10~\um/ wavelength) laser plasma interactions in the creation of superponderomotive elections. The simulations support three major findings. The first major conclusion is that the backwards acceleration of elections in the sub-relativisitic regime is more efficient when the scale length is longer than the wavelength, expressed by the condition that the ratio of the scale length to the wavelength $L/\lambda \gtrsim 1$. Second, the onset of plasma phenomenon scales with the $a_0$ value of the incident pulse, which of course takes contributions from both the intensity of the incident light and its wavelength. Finally, the physical scale of plasma phenomena scales with the wavelength, facilitating the experimental observation of features such as ponderomotive steepening.

The importance of the condition $L/\lambda \gtrsim 1$ is found from comparing the energies of near-IR and mid-IR simulations with similar $a_0$ values and scale lengths determined by $L = 1.92 \lambda$, shown on the right panel of Fig.~\ref{fig:energies} to simulations with similar $a_0$ values on the left panel with a fixed scale length $L=1.5$~\um/. In the sub-relativistic regime ($a_0 < 1 $), we find that the ``scaled'' scale length simulations with $L = 1.92 \lambda$ produce ejected electrons that happen to lie well above expectations from \cite{Wilks_etal1992} and that they also exceed that estimate by a greater amount than the corresponding fixed $L=1.5$~\um/ simulations (with similar $a_0$ value lasers) exceed that estimate. Moreover, from examining just the simulations with a fixed $L=1.5$~\um/, one finds that longer wavelength simulations produce less energetic electrons than shorter wavelength simulations. The longer wavelength cases represent values of $L/\lambda = 0.5, 0.15$ for $\lambda = $3~\um/,10~\um/, respectively. This further reinforces the idea that the $L/\lambda$ ratio needs to be unity or greater for back-directed electron acceleration to be effective in this $a_0$ regime.

The wilks scaling estimate is linear in $a_0$ in the regime where $a_0 > 1$ and quadratic in $a_0$ in the non-relativistic case ($a_0 < 1$), as discussed earlier. The scaling of energies for backwards accelerated elections for the simulations with an extended pre-plasma $L/\lambda = 1.92$  appears to have a linear power law that extends below the non-relativistic limit at $a_0 \sim 1$ and thus exceed the classical ponderomotive scaling in the sub-relativistic regime, demonstrated by the bifurcation of the Wilks scaling and the trend for backwards-accelerated elections at $a_0 \sim 1$ on the right panel of Fig.~\ref{fig:energies}. As mentioned in the caption of that figure, the lowest $a_0$ simulation did not have accelerated electrons that reached the edge of the simulation space by the end of the simulation, so it is doubtful that this power law extends to $a_0$ values well below $10^{-1}$, but this lower energy regime will be the focus of future work.

Regarding the ejected electron energy and angle spectra shown in Fig.~\ref{fig:midIR}, the analytic model actually corresponds more closely to the 1.5~\um/ scale length results for the mid-IR simulations (bottom panel) than the $L = 1.92 \lambda$ results (upper panel) which are quite a bit more energetic than expected from the model. A careful look at the upper and lower panels of Fig.~\ref{fig:midIR} plot shows that there is a larger total number of ejected electrons as well for the $L = 1.92 \lambda$ simulations. As explained in \cite{Ngirmang_etal2016}, this model is purely electromagnetic, without considering plasma effects which become more important as the ratio $L/\lambda$ increases.

The importance of the ratio $L/\lambda$ could be due to a number of factors. It is well known that intense laser interactions are highly sensitive to the assumed scale length; from a physics perspective \cite[c.f.][]{gibbon} one expects that if $L/\lambda \ll 1$, then the laser will reflect off a sharp interface (much like a mirror) and only accelerate electrons that reside on the surface of the target. If instead $L / \lambda \gtrsim 1$, then the laser will interact with a more extended region of near-critical plasma that provides a more suitable environment for accelerating large numbers of electrons in the \emph{back} direction, as we consider here. It should also be emphasized that the ``scaled'' scale length  targets are significantly larger than the $L = 1.5$~\um/ targets and this extended pre-plasma can decrease the electrostatic potential at the edge of the target where the electrons are ejected. A large pre-plasma layer can also provide a larger return current for escaping elections. These are possible explanation for both the increased energy for the simulations where $L=1.92\lambda$ as well as for the larger number of ejected electrons, as discussed earlier and shown in Fig.~\ref{fig:midIR}.  Finally, the ratio $L/\lambda$ is relevant especially in the sub-relativistic regime ($a_0 < 1$), and as discussed in Section~\ref{subsec:reflection} and Fig.~\ref{fig:reflection}, different plasma phenomena is observed different across $a_0$ regimes. These effects, as well as the confluence of the $a_0$ value with the ratio $L/\lambda$, will be investigated in future work.

Finally, as discussed in the previous section and shown in Fig.~\ref{fig:ponderomotive}, we anticipated the scaling of plasma features with the indicident laser wavelength, and observed ponderomotive steepening in the longer wavelength IR simulations with $a_0 \sim 1$ with multiple peaks in the density distribution. While it is out of the scope of this paper to design in detail a mid-IR experiment that would create these conditions and convincingly detect these density modulations using interferometry, the result is very encouraging from an experimental perspective. We comment here to say (1) that high-resolution interferometric systems have been demonstrated with soft X-ray wavelengths and used productively for research \cite[e.g.][]{Grava_etal2008} and (2) high repetition rate laser experiments with rapidly recovering, highly-reproducible liquid targets can potentially be used with a high acquisition rate interferometric system \cite[e.g.][]{Feister_etal2014} to study how these features develop over time. To be clear, this is not a comment on a particular laser system, but rather an invitation for experimental groups to consider the problem. It remains beyond the scope of this paper to determine how capable the the high repetition rate 3~\um/ wavelength ultra-intense laser system recently purchased by AFRL would be for detecting these density modulations when used with the existing 100~Hz acquisition rate interferometric system \cite{Feister_etal2014} and coupled to the existing liquid target setup there.

\section{Conclusions}
\label{sec:summary}
 In anticipation of future experiments utilizing ultra-intense, mid-infrared laser pulses and their interaction with dense targets, we used LSP 2D(3v) simulations to explore these interactions over a range of intensities and wavelengths. Similar to earlier investigations with near-IR light \cite{Orban_etal2015,Ngirmang_etal2016}, we find that intense longer IR wavelength interactions also produce highly superponderomotive electrons. Moreover, the acceleration is much more effective when the pre-plasma scale length is in similar scale to the laser wavelength, or longer. In some cases the typical ejected electron energies exceed ponderomotive expectations by orders of magnitude.
 
 The longer IR simulations also indicate that ponderomotive steepening should occur in experiments of this kind when $a_0 \sim 1$ and the pre-plasma scale length is again similar in scale to the laser wavelength. This likewise extends earlier results in the near-IR \cite{Orban_etal2015} where this phenomena was noticed in simulations. Importantly, ponderomotive steepening can create multiple peaks and valleys in the ion and electron density profile in the pre-plasma that are well known to be spaced by $\lambda/2$ peak-to-peak. To our knowledge these density modulations have never been observed experimentally. Intense longer IR laser systems coupled with high-resolution interferometry techniques should provide a promising venue for demonstrating this basic laser-plasma interaction process.
 
 


\begin{acknowledgments}


This research was sponsored by the Air Force Office of Scientific Research (AFOSR) through program managers Dr. Enrique Parra and Dr. Jean-Luc Cambier. The authors acknowledge significant support from the Department of Defense High Performance Computing Modernization Program (DOD HPCMP) Internship Program and the AFOSR summer faculty program. Supercomputer time was used on the DOD HPC Armstrong and Garnet supercomputers. The authors would also like to thank The Ohio State Department of Physics Information Technology support, specifically, Keith A. Stewart.

\end{acknowledgments}



\bibliography{ms_PoP}
\bibliographystyle{apsrev}
\appendix
\onecolumngrid

\section{Exhaustive list of simulations}
\label{ap:exhaustive}
The following is an exhaustive list of the 2D(3$v$) PIC simulations presented in this paper.
\begin{table*}
    \begin{tabular}{| x{0.7in} | x{0.7in} |
                      x{0.5in}  | x{0.5in}  | 
                      x{0.7in}  | x{0.5in}  | 
                      x{0.5in}  | x{0.7in}  |
                      x{0.7in}  | x{0.7in}  |}
    
    \hline
    Wavelength (\um/) & Intensity (W/cm$^2$) & $a_0$ & Gaussian Radius (\um/) & Pulse FWHM (fs) & Laser Energy (J) & Pre-Plasma Scale  (\um/) & Target Density (cm$^{-3}$) & Simulation Timestep (fs) & Simulation Spatial Resolution (\um/) \\ \hline
    10 & 1.64$\cdot$ 10$^{13}$ & 0.0346 & 27.5 &
    513 & 10$^{-4}$ & 1.5 & 10$^{23}$  & 0.5 & 0.25 \\ \hline
    10 & 1.64$\cdot$ 10$^{13}$ & 0.0346 & 27.5 &
    513 & 10$^{-4}$ & 19.2 & 10$^{21}$ & 0.5 & 0.25 \\ \hline
    3 & 6.08$\cdot$ 10$^{14}$ & 0.0632 &  8.25 &
    154 & 10$^{-4}$ & 1.5 & 10$^{23}$ & 0.15 & 0.1  \\ \hline
    3 & 6.08$\cdot$ 10$^{14}$ & 0.0632 &  8.25 &
    154 & 10$^{-4}$ & 5.77 & 10$^{23}$ & 0.15 & 0.1 \\ \hline
    10 & 1.64$\cdot$ 10$^{14}$ & 0.109  & 27.5 &
    513 & 10$^{-3}$ & 1.5 & 10$^{23}$  & 0.5 & 0.25 \\ \hline
    10 & 1.64$\cdot$ 10$^{14}$ & 0.109  & 27.5 &
    513 & 10$^{-3}$ & 19.2 & 10$^{21}$ & 0.5 & 0.25 \\ \hline
    0.78 & 3.46$\cdot$ 10$^{16}$ & 0.124  & 2.15 &
    40 & 10$^{-4}$ & 1.5 & 10$^{23}$ & 0.1 & 0.033\\ \hline
    3 & 6.08$\cdot$ 10$^{15}$ & 0.200 &  8.25 &
    154 & 10$^{-3}$ & 1.5 & 10$^{23}$ & 0.15 & 0.1 \\ \hline
    3 & 6.08$\cdot$ 10$^{15}$ & 0.200 &  8.25 &
    154 & 10$^{-3}$ & 5.77 & 10$^{23}$ & 0.15 & 0.1 \\ \hline
    10 & 1.64$\cdot$ 10$^{15}$ & 0.346  & 27.5 &
    513 & 10$^{-2}$ & 1.5 & 10$^{23}$ & 0.5 & 0.25\\ \hline
    10 & 1.64$\cdot$ 10$^{15}$ & 0.346  & 27.5 &
    513 & 10$^{-2}$ & 19.2 & 10$^{21}$ & 0.5 & 0.25\\ \hline
    0.78 & 3.46$\cdot$ 10$^{17}$ & 0.392  & 2.15 &
    40 & 10$^{-3}$ & 1.5 & 10$^{23}$ & 0.1 & 0.033 \\ \hline
    3 & 6.08$\cdot$ 10$^{16}$ & 0.632 &  8.25 &
    154 & 10$^{-2}$ & 1.5 & 10$^{23}$ & 0.15 & 0.1 \\ \hline
    3 & 6.08$\cdot$ 10$^{16}$ & 0.632 &  8.25 &
    154 & 10$^{-2}$ & 5.77 & 10$^{23}$ & 0.15 & 0.1 \\ \hline
    10 & 1.64$\cdot$ 10$^{16}$ & 1.09  & 27.5 &
    513 & 10$^{-1}$ & 1.5 & 10$^{23}$ & 0.5 & 0.25\\ \hline
    10 & 1.64$\cdot$ 10$^{16}$ & 1.09  & 27.5 &
    513 & 10$^{-1}$ & 19.2 & 10$^{21}$ & 0.5 & 0.25\\ \hline
    0.78 & 3.46$\cdot$ 10$^{18}$ & 1.24  & 2.15 &
    40 & 10$^{-2}$ & 1.5 & 10$^{23}$ & 0.1 & 0.033 \\ \hline
    3 & 6.08$\cdot$ 10$^{17}$ & 2.00 &  8.25 &
    154 & 10$^{-1}$ & 1.5 & 10$^{23}$ & 0.15 & 0.1 \\ \hline
    3 & 6.08$\cdot$ 10$^{17}$ & 2.00 &  8.25 &
    154 & 10$^{-1}$ & 5.77 & 10$^{23}$ & 0.15 & 0.1 \\ \hline
    10 & 1.64$\cdot$ 10$^{17}$ & 3.46  & 27.5 &
    513 & 1 & 1.5 & 10$^{23}$ & 0.5 & 0.25 \\ \hline
    10 & 1.64$\cdot$ 10$^{17}$ & 3.46  & 27.5 &
    513 & 1 & 19.2 & 10$^{21}$ & 0.5 & 0.25\\ \hline
    0.78 & 3.46$\cdot$ 10$^{19}$ & 3.92  & 2.15 &
    40 & 10$^{-1}$ & 1.5 & 10$^{23}$ & 0.1 & 0.033 \\ \hline
    3 & 6.08$\cdot$ 10$^{18}$ & 6.32 &  8.25 &
    154 & 1 & 1.5 & 10$^{23}$ & 0.15 & 0.1\\ \hline
    3 & 6.08$\cdot$ 10$^{18}$ & 6.32 &  8.25 &
    154 & 1 & 5.77 & 10$^{23}$ & 0.15 & 0.1\\ \hline
    10 & 1.64$\cdot$ 10$^{18}$ & 10.9  & 27.5 &
    513 & 10& 1.5 & 10$^{23}$ & 0.5 & 0.25 \\ \hline
    10 & 1.64$\cdot$ 10$^{18}$ & 10.9  & 27.5 &
    513 & 10 & 19.2 & 10$^{21}$ & 0.5 & 0.25\\ \hline
    0.78 & 3.46$\cdot$ 10$^{20}$ & 12.4  & 2.15 &
    40 & 1 & 1.5 & 10$^{23}$ & 0.1 & 0.033 \\ \hline
    3 & 6.08$\cdot$ 10$^{19}$ & 20.0 &  8.25 &
    154 & 10 & 1.5 & 10$^{23}$ & 0.15 & 0.1\\ \hline
    3 & 6.08$\cdot$ 10$^{19}$ & 20.0 &  8.25 &
    154 & 10 & 5.77 & 10$^{23}$ & 0.15 & 0.1\\ \hline
    0.78 & 3.46$\cdot$ 10$^{20}$ & 39.2  & 2.15 &
    40 & 10 & 1.5 & 10$^{23}$ & 0.1 & 0.033 \\ \hline
  \end{tabular}
    \caption{Summary of the parameters of all the \twod/ LSP PIC simulation performed sorted according to the value of $a_0$ for each simulation. The different laser parameters were chosen to range from 0.1 mJ to 10 J in laser energy for each of the three laser wavelengths, 780~nm, 3~\um/, and ~10\um/. A number of simulations were performed with a $L = 1.5$~\um/ scale length target. Other simulations were performed with $L = 1.92 \lambda$.}
\label{table:allparams}
\end{table*}

\end{document}